# Structural characterization of APPJ treated and Bismaleimide coatings and heat treated Titania-BMI

S.Shrinidhi[1], S. Suman[1], A.Shah[1], P.Prabhakar[1], A. Chaurasia[1]
A. Kumar[1], K. G. Chauhan[1], A.S.Bhattacharyya[1, 2*]

[1]Centre for Nanotechnology and [2]Centre of Excellence in Green and Efficient Energy Technology (CoE-GEET)
Central University of Jharkhand, Ranchi – 835 205, India,

*e-mail: 2006asb@gmail.com ; arnab.bhattacharya@cuj.ac.in

Bismaleimide (BMI) are thermosetting polymers mainly used in aerospace applications having properties of dimensional stability, low shrinkage, chemical resistance, fire resistance, good mechanical properties and high resistance against various solvents, acids, and water. BMI is commercially available as Homide 250. BMI coating has also been used for the corrosion protection. Metallization (AL) of BMI using vacuum evaporation was done which serves the purpose of prevention of space charge accumulation in aircraft bodies. Addition of inorganic materials like metal oxides can influence the properties of the polymer as an inorganic-organic composite. The organic-ionorganic composites have wide applications in electronics, optics, chemistry and medicine. Titanium dioxide ($TiO_2$, Titania) has a wide range of applications starting from photocatalysis, dye-sensitized solar cells to optical coatings and electronics. A BMI-$TiO_2$ composite was prepared by chemical route. Atmospheric Plasma Jet (APPJ) using Helium gas was also treated on BMI. XRD and FTIR studies of the composite system prepared at different temperatures showed its crystalline and structural configuration.

**Introduction**

Titanium dioxide (TiO$_2$, Titania) has a wide range of applications starting from photocatalysis, dye-sensitized solar cells to optical coatings and electronics. Bismaleimide (BMI) on the other hand are thermosetting polymers having properties of dimensional stability, low shrinkage, chemical resistance, fire resistance, good mechanical properties and high resistance against various solvents, acids, and water [1, 2]. BMI is commercially available as Homide 250 [3]. BMI coating has also been used for the corrosion protection [4]. Metallization as well as surface properties of BMI are published elsewhere [5, 6]. Addition of inorganic materials like metal oxides can influence the properties of the polymer as an inorganic-organic composite. The organic-ionorganic composites have wide applications in electronics, optics, chemistry and medicine [7, 8]. The BMI-TiO$_2$ nanocomposite structurally has a shape somewhat as shown in fig1.

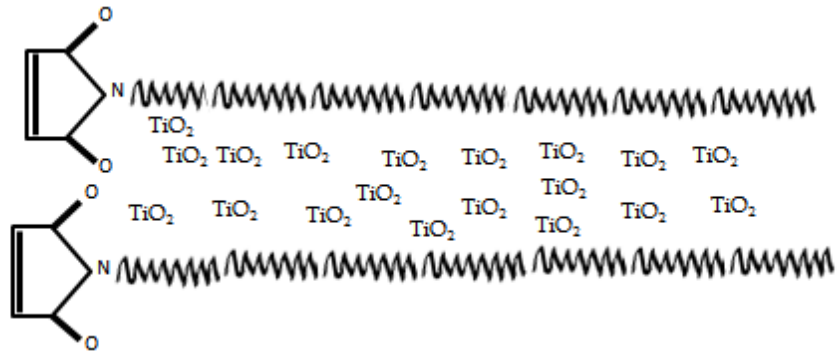

Fig 1 : A schematic representation of BMI-TiO2 composite system[7, 8]

**Experimental**

Bismaleimide (BMI) coatings were deposited on aluminium sheets by sprinkling Homide 250 BMI powders and putting the samples in oven [5]. Plasma surface modification of the polymer coatings were carried out in Argon atmosphere using Atmospheric Pressure Plasma Jet (APPJ). In Plasma surface modification (PSM), gasses dissociate in plasma and form reactive chemical species which react with the surface improving wettability, corrosion resistance, hardness and wear properties. It finds major applications in bio-medical, electronics and aerospace fields. The advantage of using APPJ lies in the fact that it overcomes the disadvantages of vacuum creation and hence large areas can be treated. The jet temperature is around 25-200 °C, charged particle density of $10^{11}$-$10^{12}$ cm$^{-3}$ and gas velocity of 12 m/s [9]. The physicochemical changes of surface of the polymer before and after plasma modification and after polymer coating were thoroughly studied.

A $TiO_2$-BMI composite system was made by taking $TiO_2$ powder, PVA and BMI (10 gm) and using magnetic stirrer. The system was quite homogenous and had no precipitation. A coating of the composite was made by uniformly depositing on a glass slide and allowing it to dry. Proto XRD at Centre of Excellence for Green Energy and Efficient Technology (CoE-GEET) CUJ was used for the XRD studies where monochromatic Cu-K$_α$ radiation having a wavelength of 1.54Å was used. Sample preparation for XRD consists of filling up the stub upto the surface in case of powder samples. For the $TiO_2$-BMI composite solution was made like a film on a glass slide as shown above and put over the stub which was subsequently transferred to the XRD chamber.

**Results and Discussions**

X-ray diffraction studies on BMI coatings on aluminum and different temperatures are shown in fig 2 which showed substrate effect of Al and mild steel. The samples BMI/Al 3 and BMI/AL 9 were bismaleimide coatings on Al showing peaks at 38°, 44°, 65° and 78° which are characteristic of aluminum substrate. The sample BMI/MS was BMI on mild steel substrate which showed peak around 44° corresponding to Fe (110). The trilayer of mild steel, BMI and AL also showed peak in the same region indicating little effect of the topmost Al layer. Fig 3 shows the effect of APPJ on the crystallinity. 1 and 2 stands for samples which have been baked in the muffle furnace starting from and initial temperature of 21°c and 200°C to 350 °C respectively. The names 1/a and 2/a indicates the samples prepared in 1 and 2 conditions but APPJ treated. Interestingly some crystallinity was observed in sample 2/a which was not coming from the substrate which has been enlarged and shown separately in Fig 4.

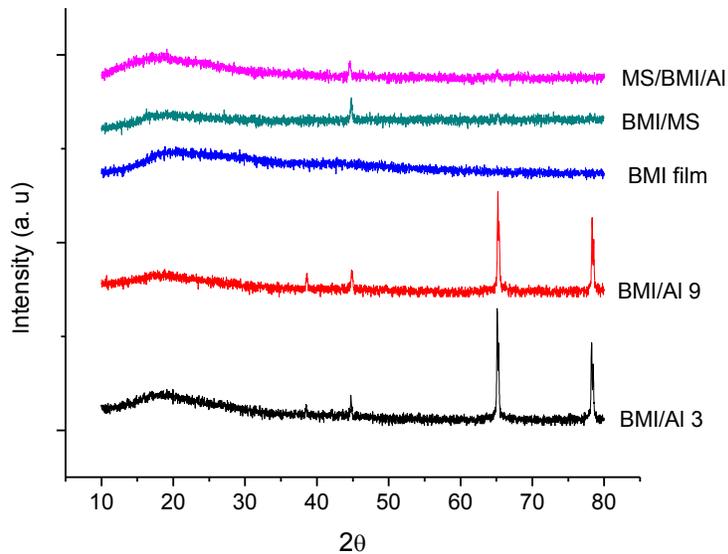

**Fig** 2: XRD of BMI coatings on different substrates and conditions

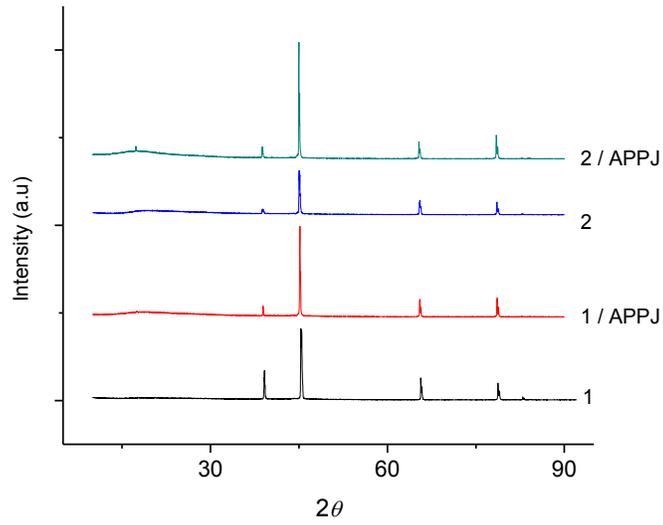

**Fig** 3: XRD of BMI coatings on Al substrates and APPJ treated

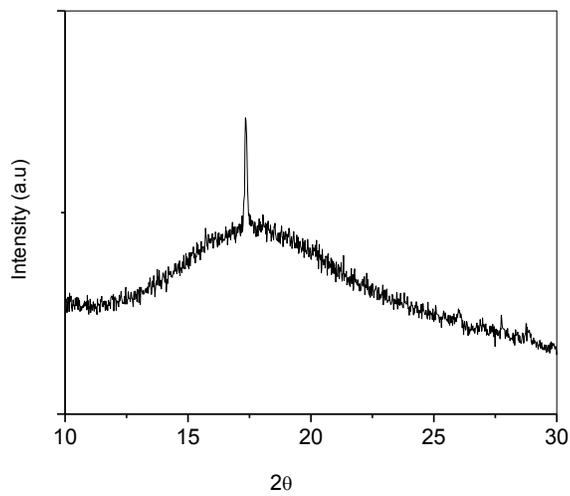

**Fig** 4: XRD showing crystallinity in APPJ treated BMI

The XRD of $TiO_2$-BMI cured at different temperatures is shown in fig 5. Shifts in the peaks were observed at 400°C indicating stress being generated due to temperature. It was also observed that peak around 19°C arising due to BMI got diminished as we increased the curing temperature and was totally absent at 400°C. The crystallinity in polymer is due to packing of molecular chains so as to produce an ordered atomic array. The sharp diffraction peaks of the indicates that it's a brittle resin [10].

The polymer molecules are only partially crystalline with crystalline regions dispersed in amorphous regions as per *fringed-micelle* model [11]. An increase in temperature results in chain disorder causing twisting and coiling of the chains preventing the strict ordering of the chain segments resulting in loss in crystallinity. The XRD pattern of the $TiO_2$-BMI sample treated at 400oC after seven days showed improved crystallinity which is probably due to the presence of $TiO_2$ which has prevented the BMI to cross link completely and as a result aided it to regain it shape on cooling. Crystallinity in polymers indicates strength and resistance to dissolution and softening by heat [12].

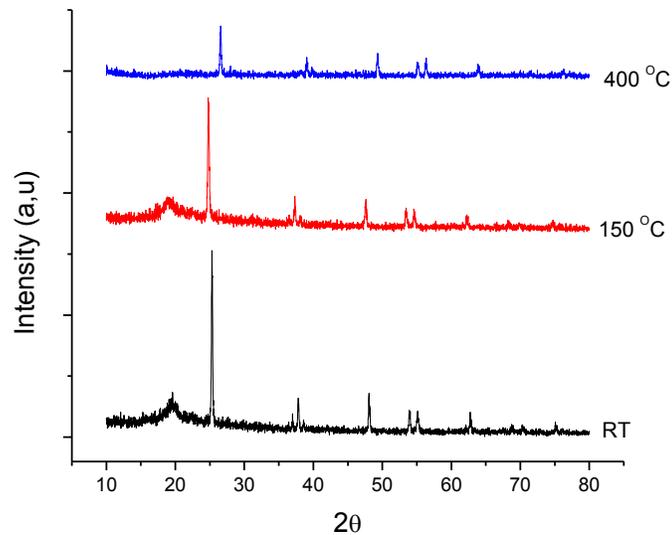

Fig 5: XRD of $TiO_2$-BMI cured at different temperatures

XRD of the $TiO_2$-BMI sample cured at 400°C was again taken after seven days to see the effect of time as shown in Fig 6. It was observed that 101 peak of anatase was much more intense compared to other phases showing a preferential crystallographic orientation of the TiO$_2$ crystallites in the composite.

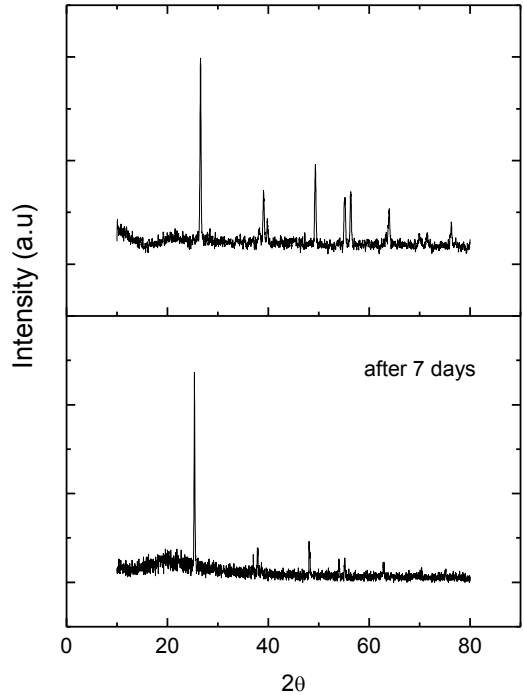

Fig 6 : XRD of TiO2-BMI cured at 400°C after 7 days

The FTIR studies are shown in fig 7. The broad peak at 3400 cm$^{-1}$ corresponds to N-H stretch. According to the literature published a band at 670 cm$^{-1}$ corresponds to TiO$_2$ (Ti-O-Ti) which is seen in the spectra of the sample at RT and which was not visible in spectra of samples at higher temperature. The peak at 830 cm$^{-1}$ at 150 °C and 400 °C corresponds to Ti-O-C. The peak at 2900 cm$^{-1}$ is due to sp$^3$ C-H stretch. The peak at 1904 cm$^{-1}$ is due to CO which is absent at 400°C. Carbonyl (C=O) groups were observed at 1710 cm$^{-1}$ which showed a shift to 1725 cm$^{-1}$ at RT indicating strengthening of H bonding due to temperature[13]. The peak at 1640 cm$^{-1}$ is due to benzene ring. The peak at 1515 cm$^{-1}$ at 150°C and 1504 cm$^{-1}$ at 400°C was due to Benzene ring. The peaks at 950 cm$^{-1}$ at higher temperature was again due to R-CH=CH-R bonds in the middle of the chain. Lu et al has however discussed about end chain bonds R-

$CH=CH_2$ at 920cm$^{-1}$ which were absent in this case [8]. The details of the different peaks are given in Table 1, 2 and 3.

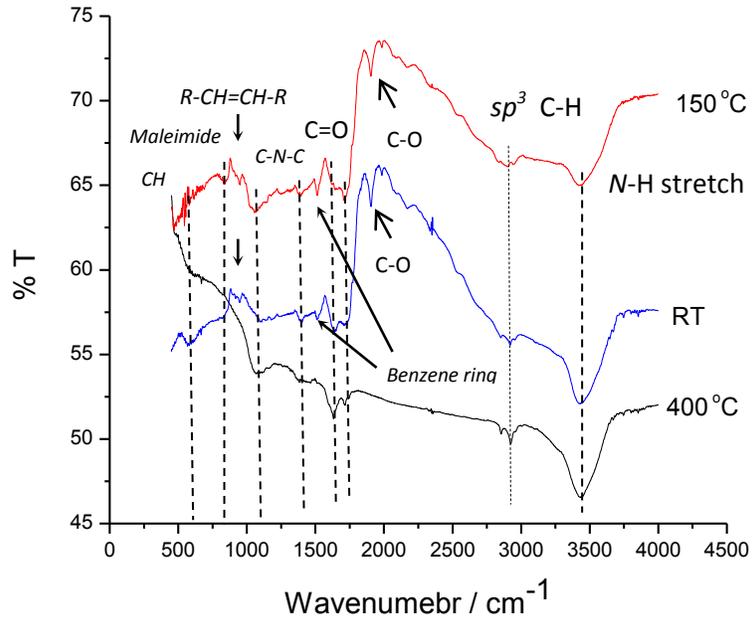

Fig 7: FTIR of TiO$_2$-BMI cured at different temperatures

Table 1: The FTIR peaks for TiO$_2$-BMI at room temperature (RT)

| Temp | Peak Positions (cm$^{-1}$) | Signature |
|---|---|---|
| RT | 3426 | N-H stretch |
|  | 2926 | C-H stretching |
|  | 1904 | C-O |
|  | 1720 | C=O symmetrical stretching |
|  | 1393 | CN and CNC stretching |
|  | 1390 | C-O stretching |
|  | 1511 | Benzene ring |

| | 949 | maleimide |
|---|---|---|
| | 580 | $TiO_2$ bonds / CH |

Table 2: The FTIR peaks for $TiO_2$-BMI at 150°C

| Temp | Peak Positions (cm$^{-1}$) | Signature |
|---|---|---|
| 150 | 3426 | N-H stretch |
| | 1904 | C-O |
| | 1710 | C=C |
| | 1720 | C=O |
| | 1509 | Benzene ring |
| | 1390 | C-O stretch |
| | 1055 | C-O stretch |
| | 833 | maleimide |
| | 950 | maleimide |
| | 1393 | CN and CNC stretching |

Table 3: The FTIR peaks for $TiO_2$-BMI at 400 °C

| Temp | Peak Positions (cm$^{-1}$) | Signature |
|---|---|---|
| 400 | 1633 | Alkenyl C=C stretch and amide C=O stretch |
| | 2930 | alkyl C-H stretch and carboxylic acid O-H stretch |
| | 3435 | N-H stretch |
| | 1055 | C-O stretching |

Raman studies of both pristine and APPJ treated samples showed no active modes (Fig 8) which indicates no polarizability change.

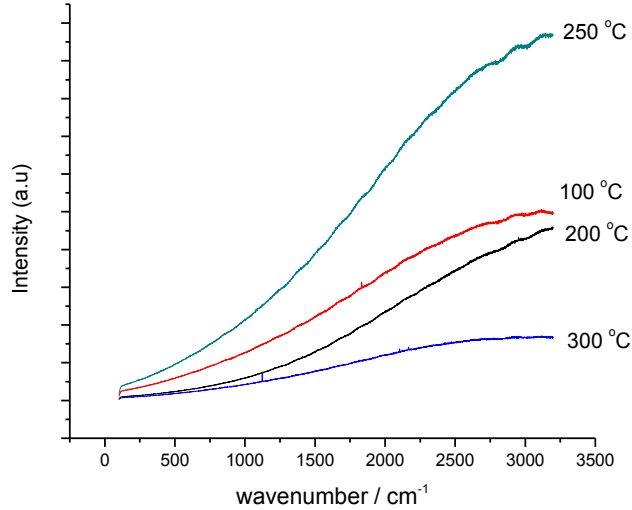

Fig 10: Raman Spectra of BMI/Al coatings at different conditions

**Summary**

Bismaleimide (BMI) coatings were deposited on aluminium sheets by a simple process of powder sprinkling and baking. The coating were well adhered to the substrate and were treated with APPJ using He gas. Crystallinity was observed in the APPJ treated BMI coatings by XRD. The samples were not Raman active. The BMI powders were blended with $TiO_2$ to form an organic-ionganic composite system. Temperature was found to have significant effect on the structure and composition of the composite system. The future scopes of work include using other gas plasma like Oxygen or Nitrogen and see the morphological as well as chemical changes taking place. Blending BMI with other materials to form composites and characterization through XPS can also be done.

**Acknowledgements**

The research was funded by DST Young Scientist Fast Track Scheme (SERB-Start up Research Grant SR/FTP/ETA-35/2011 and Centre of Excellence in Green and Efficient Energy technology (CoE-GEET) under FAST Scheme, MHRD India.